\documentclass[12pt]{emulateapj}
\journalinfo{The Astrophysical Journal Letters, 618:L000 L000, 2005 January 10}
\submitted{}
\shorttitle{Variable absorption features in  PG~1404+226}
\shortauthors{Dasgupta et al.}

\usepackage{graphicx}

\begin{document}

\title{Detection of absorption features in the
X-ray spectrum of the narrow-line Quasar PG~1404+226:
Possible evidence for accretion disk winds}

\author{Surajit Dasgupta \altaffilmark{1},
 A. R. Rao \altaffilmark{1},
 G. C. Dewangan \altaffilmark{2},
 V. K. Agrawal \altaffilmark{3} }

\altaffiltext{1}{Department of Astronomy and Astrophysics, Tata Institute
 of Fundamental Research, Mumbai-400005, India, surajit@tifr.res.in, arrao@tifr.res.in}
\altaffiltext{2}{Department of Physics, Carnegie Mellon University, Pittsburgh,
PA 15213 USA, gulabd@cmu.edu}
\altaffiltext{3}{Inter-University Center for Astronomy and Astrophysics,
Pune-411007, India, agrawal@iucaa.ernet.in}

\begin{abstract}
   We present the results of an analysis of data from {\it XMM-Newton}
   and {\it CHANDRA} observations of the high luminosity narrow-line
   quasar PG~1404+226. We confirm a strong soft X-ray excess in the X-ray
   spectrum and we find rapid variability (a factor of two in about 5000
   s). When the X-ray spectrum is fit with a two component model which
   includes a power-law and a blackbody component, we find that low
   energy absorption lines are required to fit the data. If we interpret
   these lines as due to highly ionized species of heavy elements in an
   outflowing accretion disk wind, an outflow velocity of $\sim$ 26000
   km s$^{-1}$ could be derived. One interesting feature of the present
   observation is the possible detection of variability in the absorption
   features: the absorption lines are visible only when the source is bright.
   From the upper limits of the equivalent widths (EW) of the absorption
   lines during the low flux states and also from the model independent
   pulse height ratios, we argue that the strength of absorption is lower
   during the low flux states. This constraints the physical size of the
   absorbing medium within 100 Schwartzschild radius ($R_g$) of the
   putative supermassive black hole. We also find a marginal evidence for
   a correlation between the strength of the absorption line and the X-ray
   luminosity. 

\end{abstract}


	\section{Introduction}
    \par Narrow-line Seyfert 1 galaxies (NLS1s) have very remarkable X-ray
    properties: they show evidence for strong excess of soft X-rays (dominant
    below $\sim$2 keV) above the hard X-ray continuum extrapolation and rapid
	X-ray variability \citep{bbf96}. Complex absorption features are also
	common in many of these sources. Recently \citet{pou03a,pou03b} detected
	several absorption lines in the narrow emission line quasars PG~1211+143
	and PG~0844+349 which are blue-shifted indicating relativistic outflow
	of highly ionized material. They suggest that these outflows form a
	significant component in the mass and energy budgets of systems accreting
	at or above the Eddington rate \citep{kp03}.
	
	\par PG~1404+226 (V=15, M=-23.4, z=0.098) is one of the most extreme
	narrow line Fe~II quasar. Its $H_\beta$ FWHM is 880~km s$^{-1}$ \citep{bg92}.
	The {\it ROSAT} spectrum (0.1 - 2 keV) of PG~1404+226 is steep
	($\Gamma\sim 3$) and shows rapid (factor of 2 in 10 hours) flux 
	variability. Moreover flux selected spectral analysis revealed the 
	presence of an absorption edge around 0.8 - 1 keV whose energy shift to
	higher value when the source brightens \citep{um96}. The X-ray spectrum
	of PG~1404+226 was variable during an {\it ASCA} observation \citep[hereafter
	U99]{com97,ulr99} and characterized by a strong soft excess below 2 keV
	whose luminosity in 0.4 - 2 keV band ($7\times 10^{43}$ ergs s$^{-1}$)
	was a factor of 3 greater than 2 - 10 keV luminosity. \citet{com97} have
	also found an absorption edge at 1.07 keV and they suggested an over
	abundance of iron. \citet{lei97} compared the absorption feature found in
	{\it ASCA} spectrum with absorption by ionized oxygen and derived a high
	velocity of ionized outflow from that object.
	
	\par In this {\it Letter} we present the {\it XMM-Newton} and {\it CHANDRA}
	observations of PG~1404+226. We detect absorption features in the {\it 
	XMM-Newton} data and find some evidence for such lines in the {\it CHANDRA}
	data. We have reanalyzed {\it ASCA} data which corroborates this conclusion.
	The source shows high variability in all the three observations and most
	remarkably, the line features show variability in time scales of $\sim$
	5000 s, giving a direct dynamical size constraint of 100 $R_g$ for the
	absorbing medium.

	\begin{figure*}
	\centering
	\includegraphics[scale=0.75]{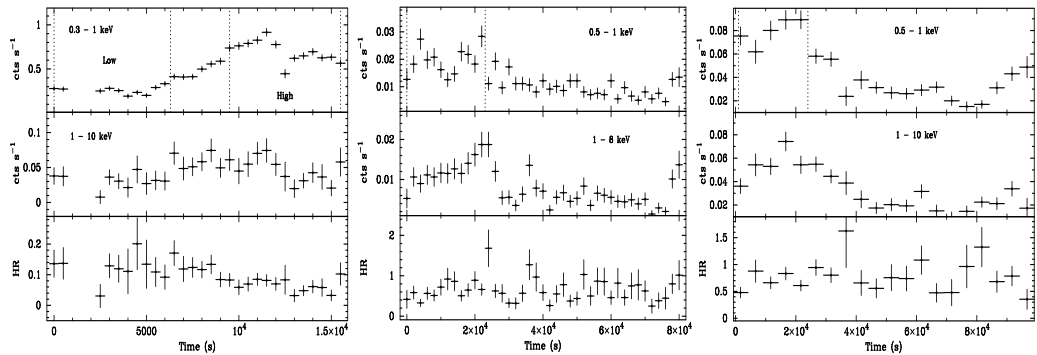}
	\caption{(a) EPIC-PN 500s bin (left), (b) CHANDRA-ACIS 2000s bin
	(middle), and (c) ASCA-SIS (SIS0 AND SIS1 added) 5000s bin light
	curves and the corresponding hardness ratio (see text).
	\label{timeseries} }
	\end{figure*}

	\section{Data and Analysis}
	\par PG~1404+226 was observed by {\it XMM-Newton} on 2001 June 18 using
	the European Photon Imaging Camera (EPIC), and the reflection grating
	spectrometer (RGS) for about 21 ks. RGS data are not useful because of
	poor signal to noise ratio. The observation data files were processed and
	filtered using the same criterion as discussed in \citet{dg04}. The high
	energy particle background flaring intervals were excluded. This resulted
	in 'good' exposure time of $\sim 14 {\rm~ks}$ for the EPIC-PN. The net
	count rate is 0.66 s$^{-1}$. The Chandra High Energy Transmission Grating
	Spectrometer (HETGS) observation of this source was performed during 2000
	July 22 for a total duration of 80 ks. The spectrum is created using
	0$^{th}$ order image of the source because of low counts collected in
	HEG and MEG. PG~1404+226 was observed by {\it ASCA-GIS/SIS} on 1994 July
	13-14. In this paper SIS data are reanalyzed. The SIS was operating in
	1-CCD mode and the data were collected in Faint mode. Standard criterion
	for good time selections (U99) have been applied.
	\par EPIC-PN light curves of bin size 500 s (background subtracted) of 
	PG~1404+226 in the energy range 0.3 - 1 keV (soft), 1 - 10 keV (hard)
	are plotted in Figure~\ref{timeseries}. The energy bands were chosen to
	separate approximately the two spectral components -- a power-law and a
	soft excess component generally observed from NLS1s \citep{lei99}. It is
	evident that the X-ray emission from PG~1404+226 varied strongly during
	the {\it XMM-Newton} observation. The average count rates in the low and
	high flux states (demarcated by dotted lines in Figure~\ref{timeseries})
	are $0.26 \pm0.01$ and $0.69\pm0.02$ s$^{-1}$ respectively in soft band,
	$0.029\pm0.007$ and $0.042\pm0.007$ s$^{-1}$ respectively in hard. The
	hardness ratio (defined as the ratio between flux of hard band to that
	of soft band) is decreasing with time, implying that the spectrum softens
	in the high state. {\it ACIS} and {\it SIS} light-curves 
	(Figure~\ref{timeseries}) show similar timescale of variability.  

	\par Photon energy spectra of PG~1404+226 and associated background 
	spectra were accumulated from the EPIC-PN, ASCA-SIS and CHANDRA-ACIS
	data. The pulse invariant channels were grouped such that each bin
	contains at least 20 counts. Data above 8 keV are not used in spectral
	analysis because background counts are comparable with source counts in
	this regime. All the spectral fits were performed with the XSPEC 11.2.0
	and using the $\chi^2$ statistics. The quoted errors on the best-fit model
	parameters are at the $90\%$ confidence level ($\Delta \chi^2 = 2.7$).
	Luminosities are derived assuming isotropic emission. A value for the
	Hubble constant of $H_0=50~km~s^{-1}~Mpc^{-1}$ and a standard cosmology
	with $q_0=0$ has been adopted.

	\par
\begin{figure}[b]
	\centering
    \includegraphics[scale=0.7]{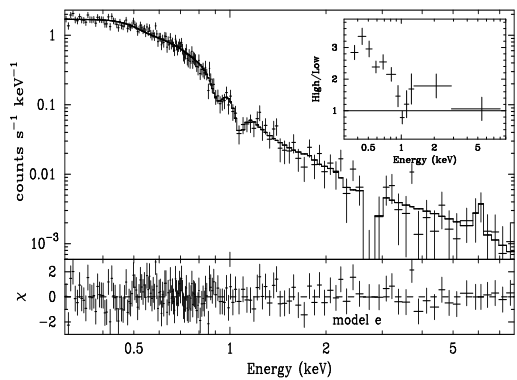}
    \caption{0.3 -- 8 keV EPIC-PN spectrum and best fitted model (model e, see
	text) and the residual spectrum (lower panel). The panel inside is 
	the PHA ratio of high and low flux states.
	\label{spec}}
    \end{figure}

\begin{figure}[h]
	\centering
    \includegraphics[scale=0.45,angle=-90]{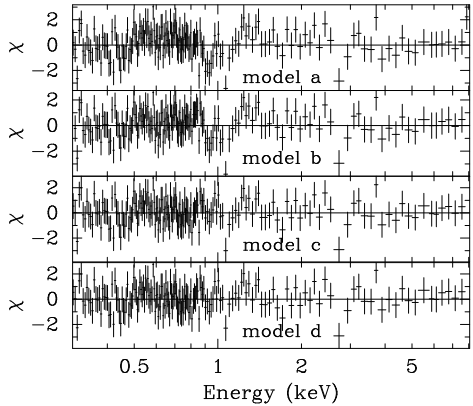}
    \caption{0.3 -- 8 keV EPIC-PN residual spectra fitted with different 
	models (see text). \label{delc}}
	\end{figure}

 We fitted a redshifted powerlaw model with a Galactic $N_H$ of
	$2\times10^{20}$ cm$^{-2}$ \citep{ewl89} to the EPIC-PN spectrum in
	the energy range of 2 - 8 keV. This provides an acceptable fit 
	($\chi^2 /dof \simeq 26/24$ and $\Gamma\sim$1.72). There	is a excess
	emission above 6.5 keV, but addition of a narrow redshifted Gaussian
	line near 6.4 keV does not improve the fit ($\Delta \chi^2_{\nu}<4$).
	Extrapolation of the best fitting 2 - 8 keV power law to 0.3 keV shows
	a huge soft excess in the spectrum. Adding a redshifted blackbody
	gives a reasonable fit ($\chi^2 /dof\sim 189/162$, $\Gamma\sim$ 1.46,
	$kT_{bb}\sim$ 108 eV). The model (model a in Figure~\ref{delc}) leaves
	significant residuals in the 0.8 - 1.2 keV (observed frame) band and
	at 3 keV. Addition of Gaussian absorption lines  at 1, 1.2, and 3 keV
	to the previous model (hereafter model e) gives a good fit with
	$\chi^2(dof) \sim134(156)$ (Figure~\ref{spec}). 
	All the derived parameter values along with the errors are given in
	the Table~\ref{tab:tab1}. We have fitted several other models to
	the data based on the previous results reported on the same source
	(Figure~\ref{delc}). An absorption component (model {\tt absori} in
	XSPEC) along with the blackbody and power-law does improve the fit
	($\chi^2 / dof \sim 174/158$ dof) but the absorption feature still
	remains (model b). Over abundance of iron (U99) cannot handle the
	absorption feature completely ($\chi^2/dof \sim 154/159$, the iron
	abundance became more than 25 times solar abundance). Adding an edge
	with the absorption model (with super iron abundance) does not improve
	the fit ($\Delta\chi^2\sim 2$). Adding an edge at $\sim$1 keV with the
	two component model (model c) can handle the absorption feature at
	1 keV ($\chi^2 /dof \sim$ 153/160) but the 1.2 keV feature still
	remains. But adding another edge (model d) for the $\sim 1.2$ keV
	feature worsen the fit ($\Delta\chi^2 \sim 1$). One edge and one line
	also cannot handle the situation. But when we adopt the model with two
	absorption lines the fit is improved drastically ($\chi^2 /dof \sim$
	144/158) and the residuals around 1 keV vanished. Adding another
	absorption line at $\sim 3$ keV improves the fit further 
	($\Delta\chi^2 \sim10$).

	\par To investigate whether the flux variability is due to changes
	in the spectral parameters we have carried out spectral analysis at low
	and high flux levels. We extracted two average spectra covering
	0 - 6.5 ks (high flux state) and 9.5 - 15.5 ks (low flux state) in
	the 0.3 - 8 keV band. The average EPIC-PN count rates are $0.26\pm0.01$
	s$^{-1}$ (low flux state) and $0.69\pm0.02$ s$^{-1}$ (high flux state).
	We fit the data using model e. The fit parameters and the observed
	fluxes in the two flux	states are given in Table~\ref{tab:tab1}. The
	observed flux in the 0.3 -- 8.0 keV range varied by a factor of 2.6
	during the high and low flux states. The blackbody flux varied by a
	factor of 3 whereas the power-law flux varied by a factor of 1.7. The
	spectral parameters are consistent with each other at 90\% confidence
	level. Since the spectral parameters and the normalizations are
	strongly coupled to each other, we can only conclude that the soft
	and hard components varied at different ratios. The absorption
	lines, on the other hand, are not required at the low flux level.
	We have included two lines (0.99, 1.17 keV) to get an estimate of
	the upper limits to the fluxes. The 90\% confidence upper limits
	on the line fluxes during the low flux state are distinctly lower
	than those found for the high flux state. The upper limit on the
	EW has an overlap with the 90\% confidence error for
	that in the high flux level. To understand the spectral variability
	in more detail, we plot the PHA ratio of high and low flux states
	with energy (Figure~\ref{spec}, inner panel). While the flux
	difference in the soft emission is remarkable, the variability
	above 1 keV is much lower. There is a clear dip at around 1 keV 
	indicating that absorption is significantly high in the high flux state.
	
	\par
\begin{figure}
    \includegraphics[scale=0.55,angle=-90]{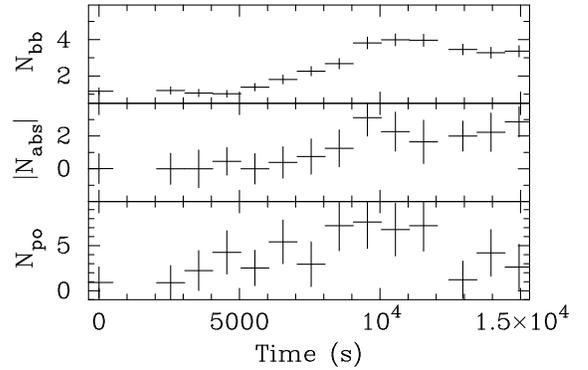}
	\centering
	\caption{Variability of normalizations (units defined in the footnote
	of Table 1) of power-law ($N_{po}$), blackbody ($N_{bb}$) and
	absorption line at 1 keV ($N_{abs}$) with time. \label{corr_var}}
    \end{figure}

 To corroborate these findings we have analysed the spectra from
	the {\it CHANDRA} and {\it ASCA} data. The detection significance
	above 1 keV in these two data sets is poor and we find a
	marginal evidence for the absorption lines in the time-averaged data.
	Results of the spectral fits to the high flux states in these two data
	sets (marked in Figure~\ref{timeseries}) are presented in 
	Table~\ref{tab:tab1}. The absorption lines are detected in the {\it
	ASCA} data ($\Delta \chi^2 \sim 10 $) and there is a marginal
	evidence for these lines in the {\it CHANDRA} data. The line energies
	and identifications in case of {\it ASCA} data are different from 
	that by \citet{lei97} probably because of different continuum modeling.
	It is clear from Table~\ref{tab:tab1} that the strength of the absoption
	lines increased when the flux increased. It is interesting to investigate
	whether these changes are correlated with each other. For this purpose we
	have divided the whole region into several parts of 1000 s duration. The
	energy channels were appropriately grouped to achieve a good signal-to-noise
	(a minimum of 5 counts per energy channel). All the spectra are fitted
	with the same model mentioned above (with C-statistics). All the
	parameters except the normalizations of blackbody, power-law and
	absorption lines are fixed to the values found from the fit result of
	the average spectrum. The relative normalization between the absorption
	lines are kept fixed. The variation of the free parameters with time is
	shown in Figure~\ref{corr_var}. It is clear that all the free parameters
	vary with time in a same fashion and the variations are similar to the
	count-rate variations. The black body flux is correlated with the line
	flux with a rank correlation coefficient of 0.79 (probability 8 $\times$
	10$^{-4}$).

\section{Discussion}
   \par The spectroscopic analysis carried in this paper using {\it XMM-Newton}
    data revealed three absorption line features at $\sim$ 1.0, 1.17, 3 keV
	with EW of 45, 60, and 364 eV, respectively. The line
	energy and EW are comparable to that found in PG~1211+143
	by \citep{pou03a}.
	We have made similar identifications 
	for the spectral lines detected in PG~1404+226 and they are given in Table
	~\ref{tab:tab2}. For some of the lines a few possible alternate line
	identifications are also given. For the two strongest lines detected by the
	EPIC-PN (0.99, 1.17 keV) we derive outflow velocities of 
	$25700^{+3300}_{-6500}$ and $43500^{+5900}_{-5900}$ km s$^{-1}$. However if we 
	consider the 1.17 keV is originated from the line Ne~IX~1s-3p (1.073 keV)
	then the measured outflow velocity will be $27000^{+5500}_{-5500}$
	km s$^{-1}$. Though there is an indication of different velocities from the
	other line identifications, we cannot rule out the possibility of all lines
	originating from similar velocity structures due to the much lower significance
	level of detection of these lines.

   \par The more striking result of our analysis is the possible detection of 
   line variability. From the model independent PHA ratio, we can see that there
   is strong spectral variability and the absorption line strength (EW) are
   also different in the two spectral states. The time scale of 
   variation ($\sim 5000s$) suggest that the absorption features originate in
   the warm absorbing material located within $\sim$ 100 R$_g$ from the central
   source (for an estimated black hole mass of 5.2 $\times$ 10$^6$ M$_\odot$).
   Since it is unlikely that the physical condition
   of the absorbing material can change with the increase in the luminosity, we
   postulate that high luminosity drives stronger winds either due to radiation
   pressure or as a result of magnetic reconnection.
   The insufficient spectral resolution of the instrument leads to large
   uncertainty of the line energy. Uncertainty in the central energy, shape,
   and possible multiple absorption components in a single feature may lead to
   incorrect velocities or identifications. But in spite of these limitations, 
   rapid variability of line strength is seen.
   \par We use the data obtained from the High Energy Astrophysics Science
   Research Archieve Center (HEASARC), provided by National Aeronautics and
   Space Administration (NASA). SD acknowledges KR scholarship of the TIFR Endowment Fund.


\begin{table*}
\footnotesize
\begin{center}
\caption{\label{tab:tab1}Best-fit spectral parameters of PG~1404+226.}
\begin{tabular}{lcccccc}
\tableline\tableline

Component & Parameter    &	\multicolumn{3}{c}{EPIC-PN$\tablenotemark{a}$}
		  				 &  \multicolumn{1}{c}{ASCA-SIS$\tablenotemark{a}$}
						 &  \multicolumn{1}{c}{CHANDRA-ACIS$\tablenotemark{a}$}\\
		  &	 	&    Average	&	High	&	Low		&	High	&   High\\
\tableline
Blackbody & $kT$ (eV)  								& $114_{-2}^{+2}$
													& $112_{-2}^{+2}$
													& $119_{-5}^{+5}$
													& $120_{-20}^{+20}$
													& $125_{-12}^{+12}$ \\
		  & $n_{bb}\tablenotemark{b}$  				& $2.97_{-0.09}^{+0.09}$
		  											& $4.34_{-0.17}^{+0.17}$
													& $1.33_{-1.00}^{+0.90}$
													& $7.46_{-2.60}^{+4.18}$
													& $2.62_{-0.37}^{+0.31}$ \\
          & $f_{bb}$(0.3-8 keV)$\tablenotemark{e}$ 	& $10.33 $
		  											& $14.82$
													& $4.86$
													& $25.82$
													& $11.05$  \\
Power law & $\Gamma$ 								& $1.59_{-0.12}^{+0.13}$
													& $1.62_{-0.16}^{+0.46}$
													& $1.27_{-0.20}^{+0.27}$
													& $1.95_{-0.42}^{+0.43}$
													& $1.55_{-0.45}^{+0.44}$ \\
		  & $n_{pl}\tablenotemark{c}$ 				& $4.42_{-0.94}^{+1.25}$
		  											& $5.05_{-1.70}^{+2.30}$
		  											& $1.97_{-0.90}^{+1.60}$
													& $37.1_{-14.5}^{+19.1}$
													& $7.23_{-3.98}^{+3.15}$  \\
          & $f_{pl}$ (0.3-8 keV)$\tablenotemark{e}$ & $2.48$
		  											& $2.75$
													& $1.55$
													& $15.32$
													& $4.26$  \\
Gaussian~1  & $E_{line}$ (keV) 						& $3.07_{-0.16}^{+0.06}$
													& $3.03_{-0.32}^{+0.10}$
													& $--$
													& $--$
													& $--$\\
          & $n_{g}\tablenotemark{d}$ 				& $-2.80_{-1.52}^{+1.56}$
		  											& $-4.00_{-3.61}^{+2.98}$
													& $--$
													& $--$
													& $--$\\
		  & $EW$ (eV)								& $-348_{-189}^{+194}$
													& $-438_{-395}^{+328}$
													& $--$
													& $--$
													& $--$\\
Gaussian~2  & $E_{line}$ (keV) 						& $1.17_{-0.02}^{+0.02}$
													& $1.17_{-0.02}^{+0.03} $
													& $1.17$
													& $1.59_{-0.08}^{+0.10}$
													& $1.16_{-1.16}^{+0.66}$ \\
          & $n_{g}\tablenotemark{d}$ 				& $-6.67_{-2.03}^{+2.13}$
		  											& $-12.43_{-3.76}^{+3.54} $
													& $>-4.80$
		  											& $-15.7_{-22.2}^{+12.3}$
		  											& $>-11.76$  \\
		  & $EW$ (eV) 								& $-60_{-18}^{+19}$
		  											& $-84_{-25}^{+24}$
													& $-22_{-47}^{+22}$
													& $-87_{-68}^{+87}$
													& $-17_{-30}^{+17}$  \\
Gaussian~3  & $E_{line}$ (keV) 						& $1.00_{-0.02}^{+0.01}$
													& $1.00_{-0.02}^{+0.01} $
													& $0.99$
													& $1.24_{-0.04}^{+0.03}$
													& $1.02_{-0.10}^{+0.10}$ \\
          & $n_{g}\tablenotemark{d}$ 				& $-13.00_{-4.03}^{+3.75}$
		  											& $-23.55_{-6.62}^{+6.61} $
													& $>-9.30$
		  											& $-35.8_{-30.0}^{+20.8}$
		  											& $-13.10_{-12.31}^{+12.09}$  \\
		  & $EW$ (eV)								& $-44_{-17}^{+17}$
		  											& $-59_{-16}^{+16}$
													& $-23_{-32}^{+23}$
													& $-81_{-68}^{+47}$
													& $-28_{-24}^{+26}$  \\
Total     & $f_{obs}$(0.3-8 keV)$\tablenotemark{e}$	& $12.44$
													& $16.93$
													& $6.52$
													& $40.25$
													& $15.04$ \\
          & $f_{int}$(0.3-8 keV)$\tablenotemark{e}$ & $15.18$
													& $20.88$
													& $7.76$
		  											& $46.58$
		  											& $17.57$ \\
	  	  & $L_{int}$(0.3-8 keV)$\tablenotemark{f}$ & $0.68$
													& $0.95$
													& $0.35$
		  											& $2.11$
		  											& $0.80$  \\
$\chi^2\tablenotemark{g}$
		  & $\chi^2/dof$	 						& $134/156$
													& $101/119$
													& $85/88$
		  											& $29/38$
		  											& $31/63$ \\
		  & $\chi^2_{1}/dof$						& $144/158$
													& $107/121$
													& $--$
													& $--$
		  											& $--$ \\
		  & $\chi^2_{12}/dof$ 						& $170/160$
													& $140/123$
													& $85/88$
		  											& $33/40$
		  											& $32/65$ \\
		  & $\chi^2_{123}/dof$	 					& $189/162$
													& $162/125$
													& $85/88$
		  											& $39/42$
		  											& $34/67$  \\
\tableline
\end{tabular}
\end{center}
\tablenotetext{a}{Parameter values for EPIC PN, ASCA-SIS and CHANDRA-ACIS
data.} 
\tablenotetext{b}{Blackbody normalization in units of $10^{-5}\times 10^{39}
{\rm~erg~s^{-1}/(d/10~{\rm~kpc})^2}$, where $d$ is the distance.}
\tablenotetext{c}{Power-law normalization in units of $10^{-5}
{\rm~photons~cm^{-2}~s^{-1}~keV^{-1}}$ at $1$ keV.}
\tablenotetext{d}{Gaussian normalization in unit of $10^{-6}
{\rm photons~cm^{-2}~s^{-1}}$}
\tablenotetext{e}{Flux in the unit of $10^{-13}{\rm~erg~cm^{-2}~s^{-1}}$}
\tablenotetext{f}{Source luminosity in the unit of $10^{44}{\rm~erg~s^{-1}}$}
\tablenotetext{g}{$\chi^2_1$: excluding Gaussian 1, $\chi^2_{12}$: excluding 
Gaussian 1 \& 2, $\chi^2_{123}$: excluding all Gaussian lines}
\end{table*}

\begin{table}
	\footnotesize
\begin{center}
\caption{\label{tab:tab2} Absorption lines identified in the parametric fit to the spectrum of PG~1404+226.}
\begin{tabular}{lcccccc}
\tableline
\tableline

$E_{source}$ & EW & $\Delta\chi^{2}/dof$ & Instrument &
$E_{lab}$ & Line & velocity \\
(keV) & (eV) &  & & (keV) &  & (km s$^{-1}$) \\

\tableline

 $1.00_{-0.02}^{+0.01}$ & $-46_{-9}^{+17} $  & 18/119   & {\it EPIC-PN}  &
 $0.921$ & Ne~IX~1s-2p 	  & $25700_{-6500}^{+3300}$  \\
 
 $1.17_{-0.02}^{+0.02}$ & $-61_{-16}^{+26}$  & 17/117   & {\it EPIC-PN}  &
 $1.022$ & Ne~X~Ly$\alpha$ & $43500_{-5900}^{+5900}$  \\
 &	&	&	& 1.073 & Ne~IX~1s-3p & $27000_{-5500}^{+5700}$ \\
  
 $1.24_{-0.04}^{+0.03}$ & $-81_{-68}^{+47}$  & 6/40     & {\it ASCA-SIS0} &
 $1.022$ & Ne~X~Ly$\alpha$ & $64000_{-11800}^{+8800}$ \\
 &	&	&	& 1.211 & Ne~X~Ly$\beta$ & $7200_{-7200}^{+7400}$ \\
  
 $1.59_{-0.08}^{+0.10}$ & $-87_{-68}^{+87}$  & 5/38     & {\it ASCA-SIS0} &
 $1.47$ & Mg~XII~L$\alpha$ & $24500_{-16000}^{+20000}$ \\
 &	&	&	& 1.211 & Ne~X~Ly$\beta$ & $94000_{-16300}^{+20500}$ \\
  
 $3.07_{-0.16}^{+0.06}$ &$-364_{-165}^{+208}$& 10/154   & {\it EPIC-PN}  &
 $2.62$ & S~XVI~Ly$\alpha$ & $51500_{-18300}^{+6900}$ \\
  
\tableline
\end{tabular}
\end{center}
\end{table}

\end{document}